\documentclass[12pt,preprint]{aastex}

\usepackage{amssymb}
\usepackage{times}
\usepackage{epsfig}
\usepackage{graphics}

\newcommand{\be}{\begin{equation}}
\newcommand{\ee}{\end{equation}}
\newcommand{\bea}{\begin{eqnarray}}
\newcommand{\eea}{\end{eqnarray}}

\slugcomment{ }

\shorttitle{Investigation of the new cataclysmic variable 1RXS \,J180834.7+101041}
\shortauthors{Yakin at al.}

\begin{document}

\title{Investigation of the new cataclysmic variable 1RXS \,J180834.7+101041
    }


\author{D. G. Yakin\altaffilmark{1}, V. F. Suleimanov\altaffilmark{2,1}, N.V.Borisov\altaffilmark{3},
V.V. Shimanskii\altaffilmark{1} and I.F. Bikmaev\altaffilmark{1,4}}
\affil{}
\email{}


\altaffiltext{1}{Dept. of Astronomy, Kazan Federal University, Kremlevskaya 18, 420008 Kazan, Russia;
screplay@mail.ru}
\altaffiltext{2}{Institute for Astronomy and Astrophysics, Kepler Center for Astro and Particle Physics,
   Eberhard Karls University, Sand 1, 72076 T\"ubingen, Germany; suleimanov@astro.uni-tuebingen.de}
\altaffiltext{3}{Special Astrophysical Observatory, Russian Academy of Sciences, Nizhnii Arkhyz, Karachai-Cherkessian Republic, 357147 Russia}
\altaffiltext{4}{Academy of Sciences of Tatarstan, ul. Baumana 20, Kazan, 420111 Russia }

\begin{abstract}
We present the results of our photometric and spectroscopic studies of the new eclipsing
cataclysmic variable star 1RXS J180834.7+101041. Its spectrum exhibits double-peaked hydrogen and
helium emission lines. The Doppler maps constructed from hydrogen lines show a nonuniform distribution
of emission in the disk similar to that observed in IP Peg. This suggests that the object can be a
cataclysmic variable with tidal density waves in the disk. We have determined the component masses
($M_{\rm WD} =0.8 \pm 0.22 M_\odot$ and $M_{\rm RD} =0.14 \pm 0.02 M_\odot$) and
the binary inclination ($i =78.^{\circ} \pm 1.^{\circ} 5$) based on
well-known relations between parameters for cataclysmic variable stars. We have modeled the binary light
curves and showed that the model of a disk with two spots is capable of explaining the main observed
features of the light curves.
\end{abstract}


\section{Introduction}

Cataclysmic variables (CVs) are currently believed
to be close binary systems at a late evolutionary stage
(Warner 1995). As a rule, CVs consist of a more
massive white dwarf (WD) (primary) and a less massive
red dwarf (secondary). The secondary overfills its
Roche lobe, causing mass transfer to the primary.
Since the matter accreting onto the WD has a
significant angular momentum, it does not fall directly
onto it but forms an accretion disk around the WD
(at weak WD magnetic fields $B \le 10^5$). At magnetic
fields $B\approx10^6-10^7$ (intermediate polars), the WD
magnetic field destroys the accretion disk at radii
smaller than the Alfven radius. Further out, accretion
occurs along WD magnetic field lines. At strong WD
magnetic fields $B\approx10^8$ (polars), the magnetic field
prevents the formation of a disk and the matter moves
along magnetic field lines from the outset. Whether
this binary will be a polar or an intermediate polar
depends not only on the WD magnetic field strength
but also on the accretion rate and the WD Roche
lobe size, i.e., on the binary period and the component
mass ratio. Clearly, short-period binaries with a low
accretion rate will be polars with a higher probability.

The secondaries of long-period CVs often have
radii exceeding appreciably those of single main sequence
stars with the same mass (Patterson et al.
2005). This is apparently a consequence of additional
heating of the secondary at the common-envelope
stage (Schreiber and G\"ansicke 2003; Shimanskii
et al. 2009). However, the secondaries manage
to relax in the time of a fairly long accretion-free
evolutionary stage in the 2-3 h period gap and their
radii become close to those of single main-sequence
stars. This allows the well-known theoretical stellar
mass-radius relation to be used, which, in turn,
makes it possible to uniquely relate the secondary
mass and the binary period (Howell et al. 2001). The
red dwarf mass-orbital period relations ($M_{\rm KK} - P_{\rm orb}$)
derived from observations confirm the theoretical
results (Patterson et al. 2005; Knigge 2006).

The accretion disk is currently believed to have a
fairly complex structure. A hot spot is formed at the
accreting gas stream-accretion disk impact site (for
a classical study of the observed properties of the hot
spot in quiescence of the dwarf nova Z Cha, see Wood
et al. 1986). It is easiest to assume that the hot spot is
the matter heated by a shock during the stream-disk
impact. However, there are numerical gasdynamic
calculations that show that the stream and the accretion
 disk are a morphologically single structure and
their interaction is impactless (see, e.g., the review
by Fridman and Bisikalo 2008). In the calculations
of this group, an extended shock emerges along the
back (relative to the direction of orbital motion) side
of the stream where the accretion disk matter collides
with it (the so-called "hot line").

Another factor breaking the axial symmetry of the
accretion disk is the tidal interaction of the secondary.
Under certain conditions, it can give rise to spiral
density waves in the disk (Boffin 2001), which were
first detected in IP Peg from observations (Steeghs
et al. 1997; Steeghs 2001; Papadaki 2008). The
calculations of the already mentioned group (Fridman
and Bisikalo 2008) showed that the tidal effect from
the secondary could give rise to a spiral density wave
consisting of one arm in the outer and partially inner
disk (for a hot accretion disk) or two arms in the
outer disk (for a cold disk). For cold accretion disks,
hydrodynamic modeling predicts the formation of a
precessional density wave in the inner disk that is
almost stationary relative to a remote observer. For
hot accretion disks, even if the mass transfer rate in
the binary changes only slightly, modeling predicts
(Bisikalo et al. 2001) the formation of a one-armed
density wave in the inner disk that revolves around
the WD approximately a factor of 5 faster than does
the secondary. This leads to emission variability with
a period of $\approx 0.2 P_{\rm orb}$.

Thus, both observations and gasdynamic calculations
suggest a complex structure of the accretion
disks; their further studies in CVs are needed for a
better understanding of this structure. The best tool
for such studies is the method of Doppler tomography
proposed by Marsh and Horne (1988). The
one-dimensional emission line profiles obtained with
a high spectral resolution during one or more complete
orbital periods serve as the input data. Using
these profiles and knowing the basic parameters of the
binary, we can reconstruct the Doppler tomogram,
which is a map of the intensity distribution for a given
emission line in velocity space I(Vx,Vy). The assumption
that a distinctive radial velocity corresponds
to the observed intensity at each point of the emission
line profile underlies the method. Thus, the line profile
at a given orbital phase is considered as a record of the
projection of the velocity field for the emitting material
onto the line of sight corresponding to this orbital
phase. Having a set of such projections (spectral line
profiles) for a set of phases covering the entire period,
we can reconstruct the intensity distribution map in
velocity space and study the spatial distribution of
the emitting plasma in terms of the adopted model
for the motion of matter (for example, by assuming
the matter to rotate in Keplerian orbits in the disk
around the white dwarf). Any nonuniformity in the
distribution of emission over the disk will be reflected
in such a tomogram.

The object 1RXS J180834.7+101041 = USNO-B1 1001-0317189 ($\alpha_{2000}=18^{h}08^{m}35^{s}.8$, $\delta_{2000}=+10^{\circ}10'30''.2$), 1RXS J1808 for short, was first detected
by the ROSAT orbital X-ray observatory as an
X-ray source and was then identified as an eclipsing
close binary system with an accreting white dwarf
(a cataclysmic variable) with a brightness of $16^m$-$17^m$
(Denisenko et al. 2008). Denisenko et al. (2008)
determined the orbital period of the binary ($P_{\rm orb} = $0.$^d$070037(1))
and detected its emission variability
with an amplitude of $\sim 1^m$ on time scales of several
weeks, which gave grounds to classify the binary as
a polar. However, a double-peaked structure of the
hydrogen and helium emission lines was detected in
the binary spectrum (Bikmaev and Sakhibullin 2008).
This suggests the existence of an accretion disk
around the white dwarf and is in conflict with the
nature of polars.

In this paper, we study in detail 1RXS J1808 based
on our analysis of new spectrometric and photometric
observations. We have investigated the brightness
distribution over the accretion disk in several emission
lines by Doppler tomography, obtained the parameters
of the radial velocity curve, and determined
the binary parameters. We have modeled the binary
light curves, which allowed the parameters of the
bright spots on the accretion disk to be refined.

\section{Observations}

\begin{figure}
\centering
\includegraphics[clip,angle=0,width=80mm]{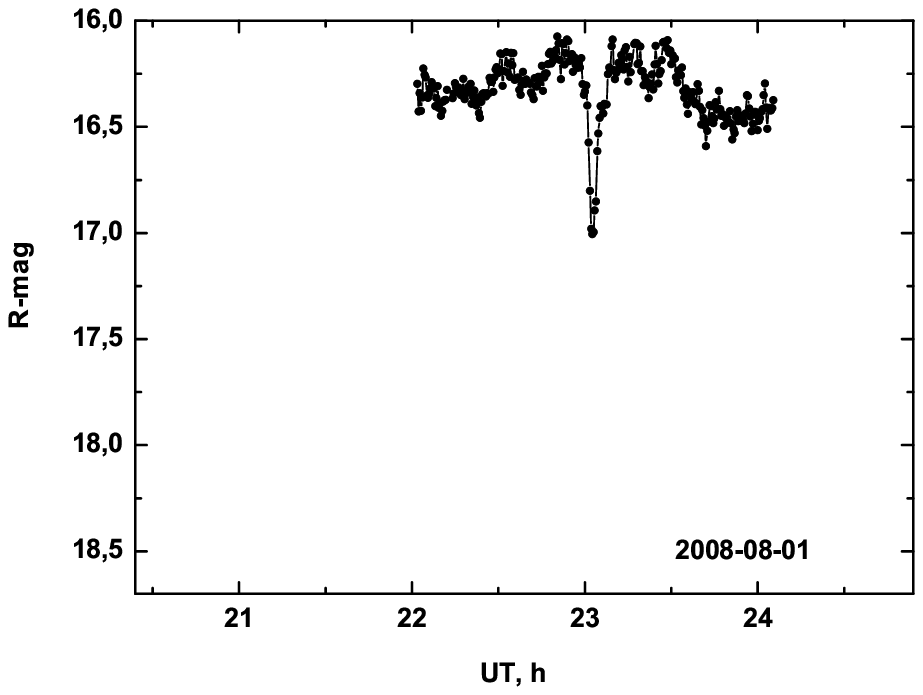}
\includegraphics[clip,angle=0,width=80mm]{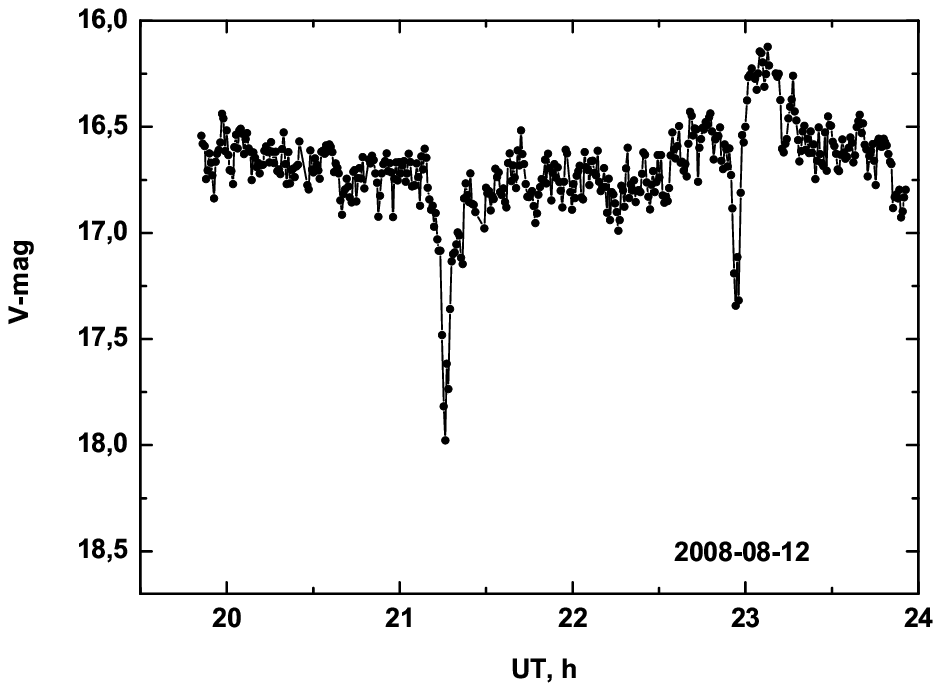}
\includegraphics[clip,angle=0,width=80mm]{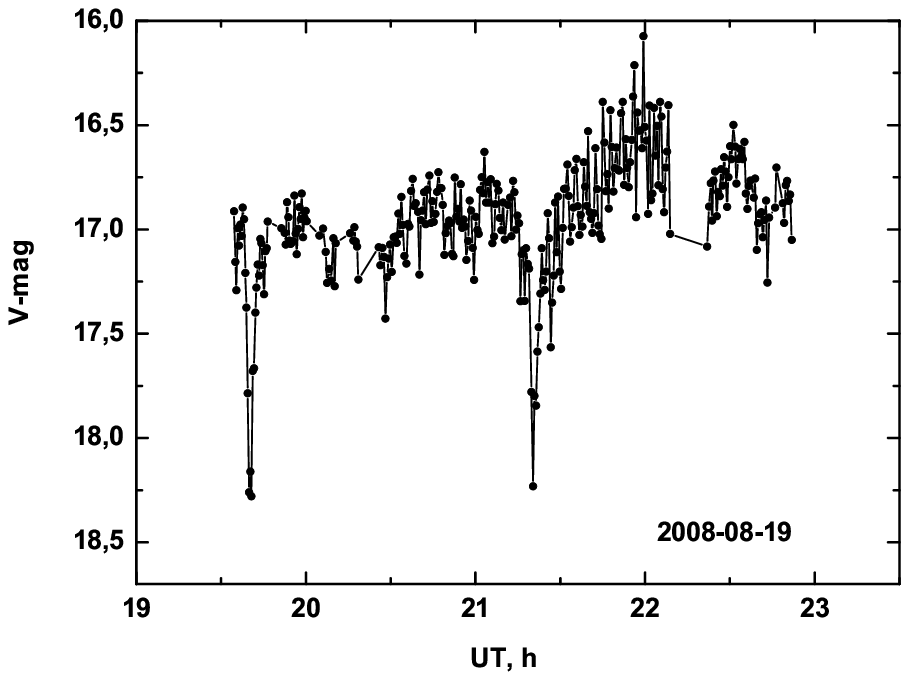}
\includegraphics[clip,angle=0,width=80mm]{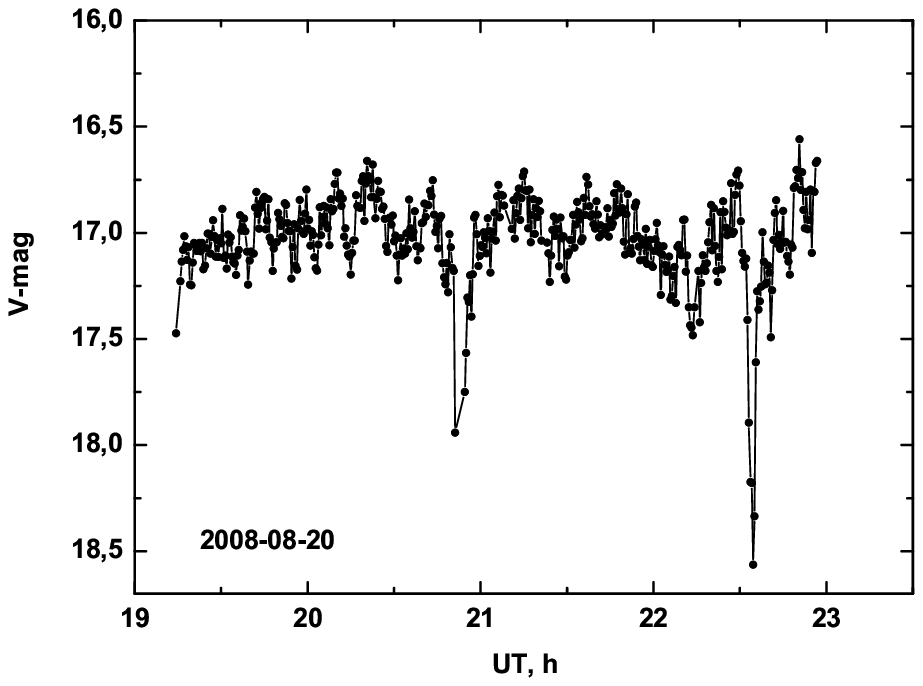}
\caption{\label{fig1}
 Observed R- and V-band light curves. }
\end{figure}

The photometric observations of 1RXS J1808
were performed with the 1.5-m Russian-Turkish
telescope RTT-150 at the TUBITAK National Observatory
 (Turkey). We used a thermoelectrically
cooled ANDOR CCD array (DW436, $2048 \times 2048$ pixels, and a pixel size of $13.5 \times 13.5$ 
$\mu$m) at a
temperature of -60$^{\circ}$C mounted at the Cassegrain
focus of the telescope. The observations were carried
out on August 1/2, 2008, in the R band with an
exposure time of 10 s and on August 12/13, 19/20,
and 20/21, 2008, in the V band with exposure times
of 15 and 20 s. The total observing time was about
13 h. Landolt's standard stars were used for the
photometric calibration. The light curves are shown
in Fig. \ref{fig1}.

\begin{figure}
\centering
\includegraphics[clip,angle=0,width=130mm]{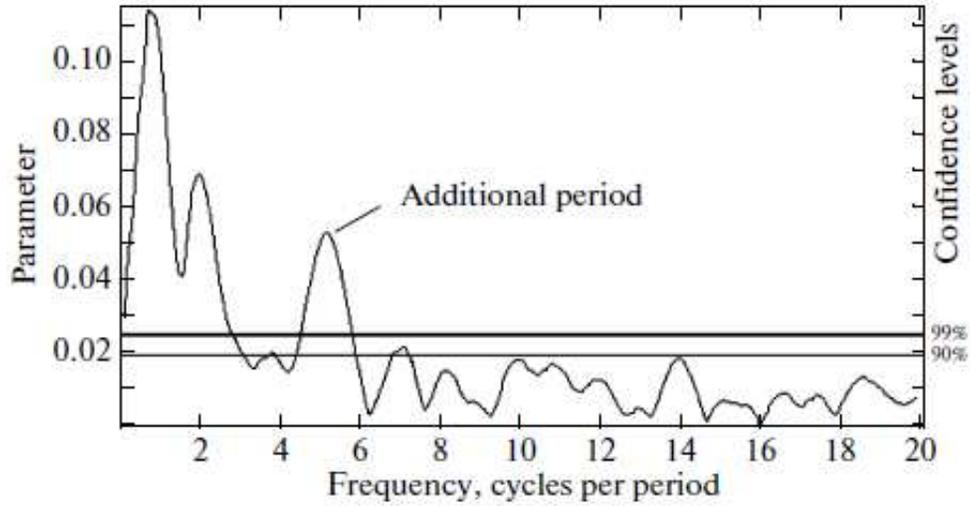}
\caption{\label{fig2}
Power spectrum for the R-band light curve. }
\end{figure}

Analysis of the light curves shows that the eclipse
depth can differ significantly from eclipse to eclipse
(see the light curve for August 12, 2008) and a general
irregular change in the binary brightness by 0$^m$.3 -- 0$^m$.5
unrelated to eclipses is observed. Some delay in
the brightening typical of binaries with an appreciable
 contribution from the hot spot is noticeable
at the brightening phase after an eclipse (see, e.g.,
 Smak 1994). The R-band light curve reveals a quasiperiodicity 
 in the binary brightness variations with a
period of $\approx 0.^d$01331
($\approx 0.19~ P_{\rm orb}$). This is confirmed by
our analysis of the power spectrum (Fig. \ref{fig2}) computed
with the EFFECT code (V. Goransky, the Sternberg
Astronomical Institute, Moscow). We failed to reveal
the corresponding brightness oscillations in the V band
 light curves because of the lower signal-to-noise
 ratio in the observational data.

\begin{figure}
\centering
\includegraphics[clip,,angle=0,width=130mm]{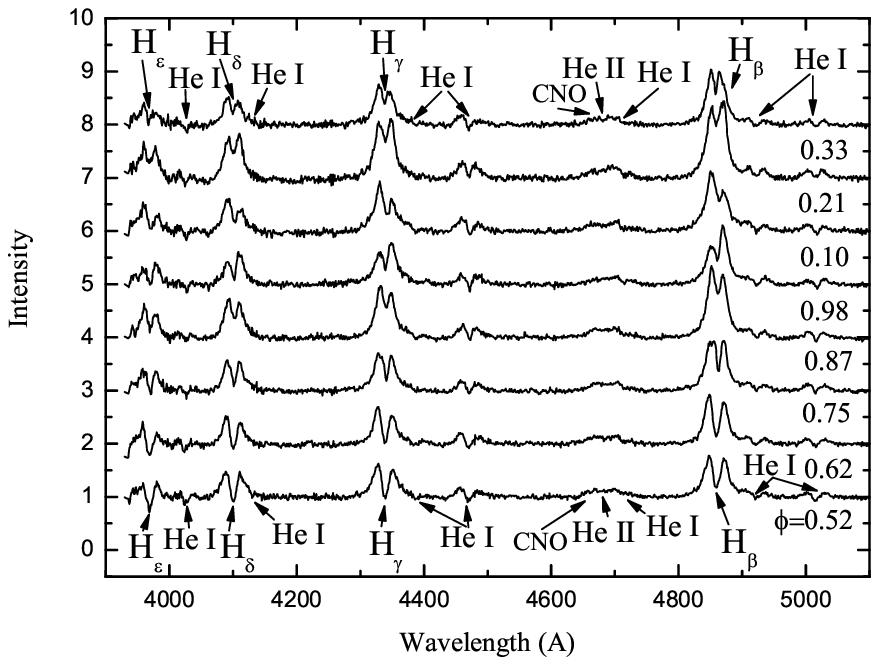}
\caption{\label{fig3}
Normalized spectra of 1RXS\,J1808 at various orbital phases (the numbers near the curves). }
\end{figure}

The spectroscopic observations of 1RXS J1808
were performed with the 6-m BTA telescope at
the Special Astrophysical Observatory of the Russian
 Academy of Sciences using the SCORPIO
focal reducer in the mode of long-slit spectroscopy
(Afanasiev et al. 2004) and an $EEV~42-40~CCD$ detector
 ($2068 \times 2048$ pixels with a pixel size of $13.5 \times 13.5 \mu$m) mounted at the prime focus. The spectra were
taken with a VPHG1200g (1200 lines/mm) prism,
which provided the spectral resolution $\Delta \lambda = 5.0$~\AA~
in the working wavelength range $\Delta\lambda$ 3950--5700~\AA.
The observations were carried out on August 9/10,
2008, under good seeing. We took
a total of 16 spectrograms with the same exposure
time of 300 s. The log of observations is presented in
Table \ref{tbl-1} and contains data on the times of observations
HJD and the orbital phases $\varphi$ calculated from the
known binary period and the eclipse time on the night
from August 1 to August 2. Consequently, the error
in the phase is 0.002
phase. The mean signal-to-noise
 ratio in the spectrograms is $S/N = 55$--$65$. For
the wavelength calibration, we simultaneously took
the spectra of an $Ar$--$Ne$--$He$-lamp. We reduced
the spectrograms according to a standard scheme
using the Scorpio2K astronomical data reduction
tools in the IDL environment. Examples of the
spectra are given in Fig.\,\ref{fig3}. The H$_{\beta}$, H$_{\gamma}$, H$_{\delta}$ and
HeI$~\lambda 4471, \lambda 5015$, HeII$~\lambda 4686$
lines are clearly seen in
the spectrum. The lines are emission ones and have a
double-peaked structure typical of accretion disks.
The interpeak minima near the line centers have
different depths at different phases. In several lines
(for example, in the neutral helium lines and H$_{\epsilon}$), the
flux at the interpeak minimum is lower than that in the
continuum at some phases. Both these facts suggest
the presence of not only emission line components but
also absorption ones in the spectrum. The relative
height of the Balmer line peaks also changes with
orbital phase. The changes near an eclipse are most
pronounced. Before phase 0.9, the relative intensity
of the blue peak is higher than that of the red one;
subsequently, the intensity of the red peak becomes
more intense during the eclipse and the flux again
becomes higher in the blue peak after the eclipse
(phases greater than 0.09).

\section{Analysis of the spectroscopic observations}

\subsection{Radial Velocity Measurement}

The radial velocities for several lines were obtained
by the Shafter (1983) method and their dependence
on the orbital phase was fitted by a sine wave. In
this case, the points near the eclipse (the phases from
-0.1 to 0.1) were disregarded. In the method used,
two narrow (compared to the width of the investigated
line) Gaussians whose centers are separated by a
distance $a$~(\AA)~ are cut out in the blue and red emission
line wings. By concurrently moving both Gaussians
over the spectrum and comparing the fluxes in the
line wings cut out by the Gaussians, it is necessary
to achieve the equality of these fluxes. The wavelength
 corresponding to half the separation between
the Gaussians can then be considered to correspond
to the wavelength of the entire investigated line and
its radial velocity can be measured. Obviously, the
result will depend on the chosen separation between
the Gaussians $a$. In an ideal case, the emission line
width is believed to be determined only by the Doppler
effect due to the Keplerian rotation of the matter in
the accretion disk, while the line intensity distribution
over the disk is axisymmetric. It can then be assumed
that the radiation in the line wings is formed closest
to the WD and it is the change in the positions of
the emission line wings that best reflects the the WD
radial velocity. Therefore, it is assumed in the method
under consideration that the larger the separation $a$,
the more accurate the WD radial velocity obtained.
At the same time, however, $a$ cannot be chosen to
be too large, because the Gaussians will begin to go
outside the emission line into the continuum. This
manifests itself in a significant increase in the relative
error $\Delta K/K$ of the amplitude of the sine wave fitting
the derived velocity curve. Therefore, the largest value
of $a$ at which the relative error of the amplitude is still
small is commonly chosen.

\begin{table}
\begin{center}
\caption{Log of observations for 1RXS J1808.\label{tbl-1}}
\begin{tabular}{|c|c|c|}
\hline
$N$ & $HJD$ & $\varphi$  \\
\hline
    & $2454000+$ &   \\
\hline
603 & 688.341 & 0.520 \\
604 & 688.345 & 0.582 \\
605 & 688.349 & 0.634 \\
606 & 688.353 & 0.697 \\
607 & 688.357 & 0.754 \\
608 & 688.361 & 0.811 \\
609 & 688.365 & 0.868 \\
610 & 688.369 & 0.924 \\
611 & 688.373 & 0.981 \\
612 & 688.377 & 0.039 \\
613 & 688.381 & 0.096 \\
614 & 688.385 & 0.154 \\
615 & 688.389 & 0.211 \\
616 & 688.393 & 0.268 \\
617 & 688.397 & 0.325 \\
618 & 688.401 & 0.382 \\
\tableline
\end{tabular}
\tablecomments{$N$ is the spectrum number, HJD is the heliocentric Julian date, $\varphi$
is the orbital phase.}
\end{center}
\end{table}

\begin{figure}
\centering
\includegraphics[clip,width=120mm]{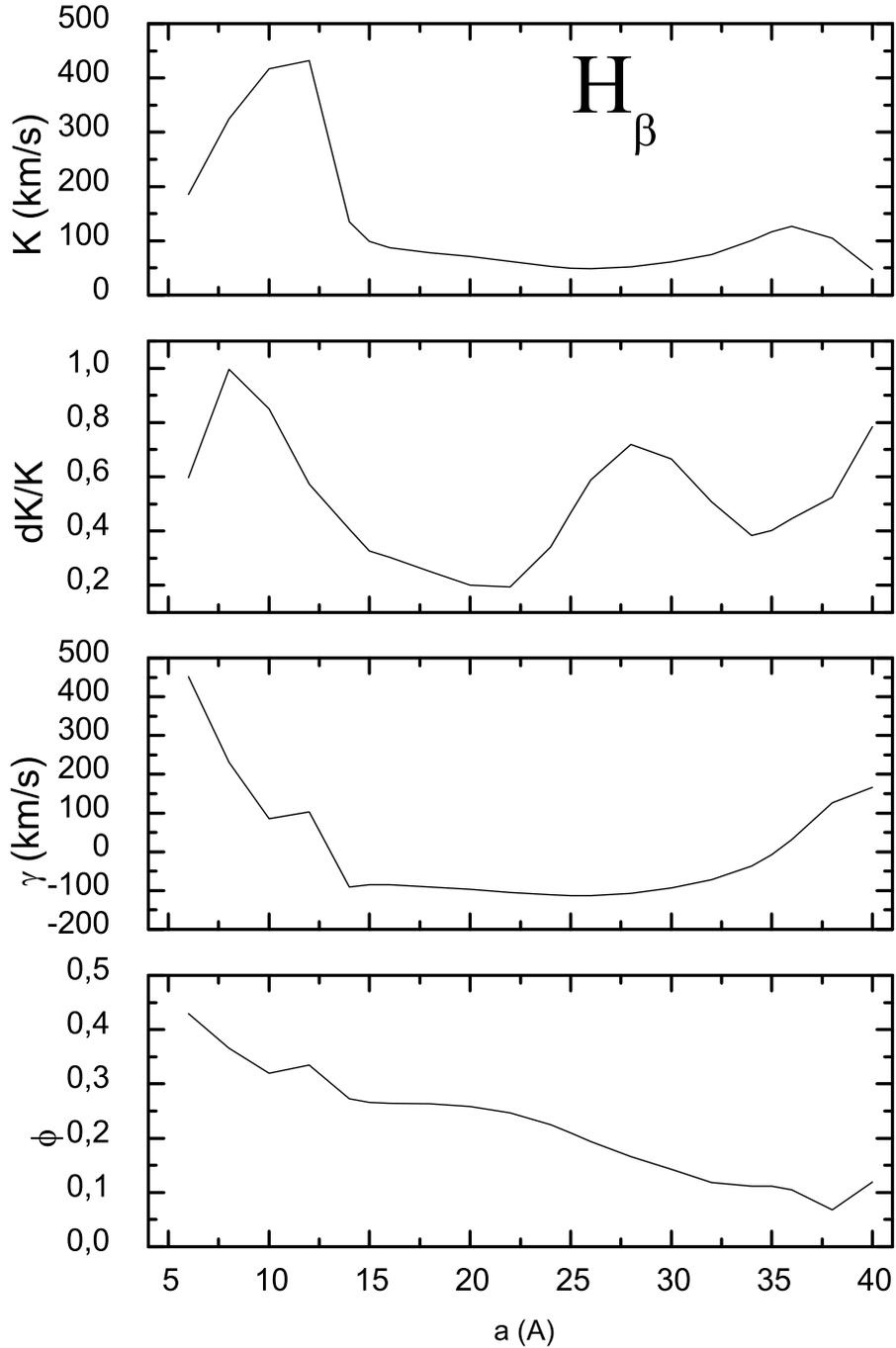}
\caption{\label{fig4}
Diagnostic diagrams of the radial velocity curves for 1RXS\,J1808 for the H$_\beta$
line. From top to bottom: the plots of
radial velocity semi-amplitude K, relative error $\Delta$K/K, gamma velocity $\gamma$, and phase shift in phase units against the separation between the sine waves $a$. The parameters obtained at $a$ =22\AA ~were chosen, because the minimum relative error $\Delta$K/K
is achieved for this separation between the sine waves.  }
\end{figure}

The corresponding diagnostic diagrams showing
the variation in the parameters of the fitting sine
waves with $a$ for H$_{\beta}$. data as an example are presented
in Fig. \ref{fig4}. As we see from the diagnostic curves,
the parameters of the sine wave change greatly for
different $a$, but the relative error in $K$ begins to increase
greatly at $a>22$\,\AA. Therefore, the parameters
corresponding to $a$=22\,\AA  ~were taken as the final ones.

\begin{figure}
\centering
\includegraphics[clip,angle=0,width=90mm]{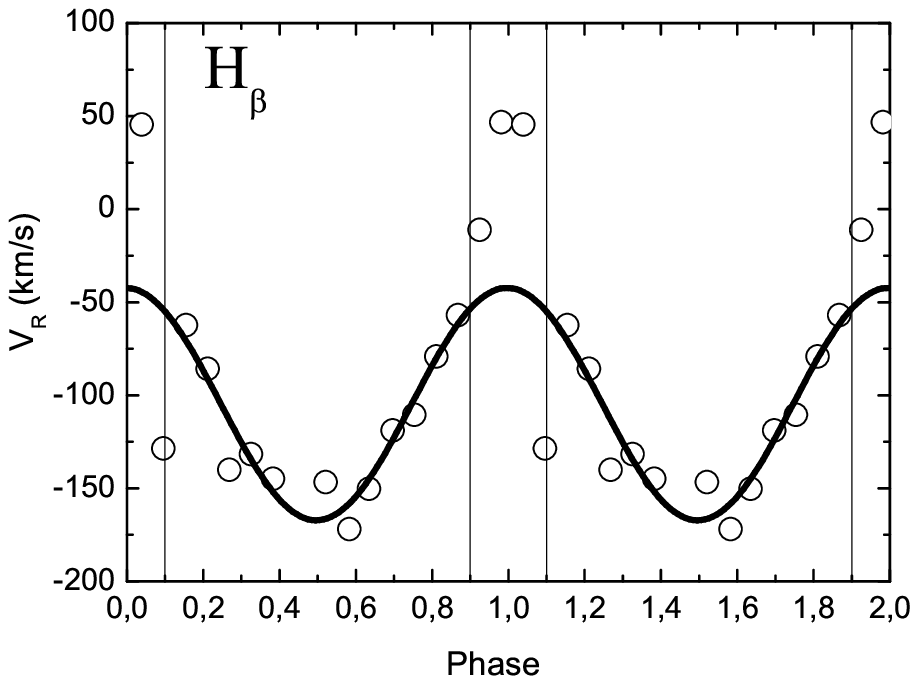}
\includegraphics[clip,angle=0,width=90mm]{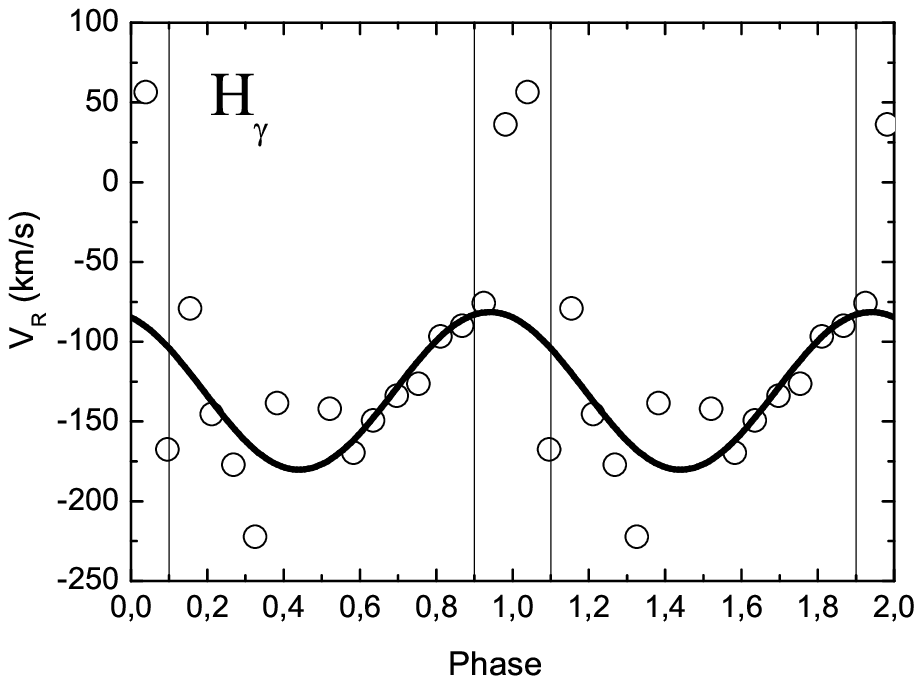}
\includegraphics[clip,angle=0,width=90mm]{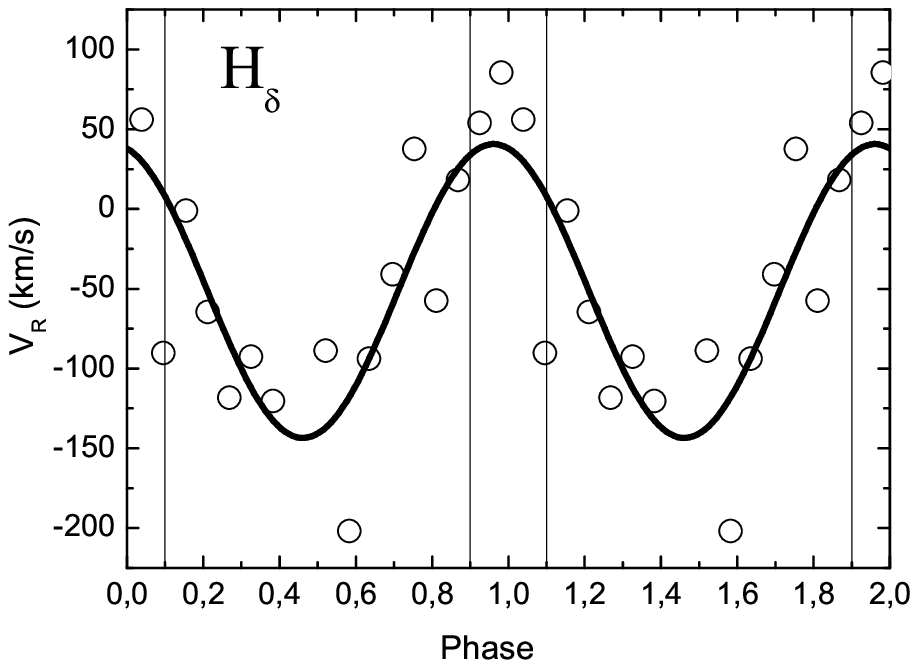}
\caption{\label{fig5}
Radial velocity curves of 1RXS\,J1808 derived from the H$_\beta$, H$_\gamma$, and H$_\delta$
lines. The vertical lines indicate the region near
the eclipse the points inside which were disregarded when fitting the data by the sine waves. }
\end{figure}

The derived radial velocity curves with their best
fits are shown in Fig. \ref{fig5} for the H$_{\beta}$, H$_{\gamma}$, H$_{\delta}$
lines. Zero phase corresponds to the eclipse. The
parameters of the fits are presented in Table \ref{tbl-2}. The
averaged amplitude is $K_1$ = 63 $\pm$ 8 km s$^{-1}$.Note
that the error in the parameters increases greatly with
Balmer line number.

Note that the velocity near the maximum of the
radial velocity curves rather than the center-of-mass
velocity, as it must be for the orbital motion, occurs
at zero phase in all cases. Therefore, we believe
that the derived radial velocity curves do not describe
the orbital motion of the white dwarf but more likely
reflect the radial velocity of the bright spots in the
disk, where these lines originate. The radial velocities
determined from helium lines are random and unsuitable
for analysis.

\begin{table}
\begin{center}
\caption{Parameters of the radial
velocity curves for 1RXS\,J1808. \label{tbl-2}}
\begin{tabular}{|c|c|c|c|}
\hline
Line &  $K$ & $\gamma$ & $\Delta \varphi_0$ \\
\hline
$H_{\beta}$ & $62\pm4$ & $-104$ & $0.25\pm0.01$ \\
$H_{\gamma}$ &$48\pm9$ & $-132$ & $0.19\pm0.02$ \\
$H_{\delta}$ & $87\pm15$ & $-53$ & $0.21\pm0.01$ \\
Mean & $63\pm8$ & $-102$ & $0.22\pm0.02$ \\
\hline
\end{tabular}
\tablecomments{$K$ is the semi-amplitude of the radial velocity curve in
km s$^{-1}$,  $\gamma$ is the center-of-mass radial velocity in km s$^{-1}$, and
$\Delta \varphi_0$ is the phase shift of the curve. Presented parameters are obtained
 at $a$= 22\AA. The minimum relative errors $\Delta K/K$ are reached at this $a$. }
\end{center}
\end{table}

\subsection{Doppler Tomography}

\begin{figure}
\centering
\includegraphics[clip,width=130mm]{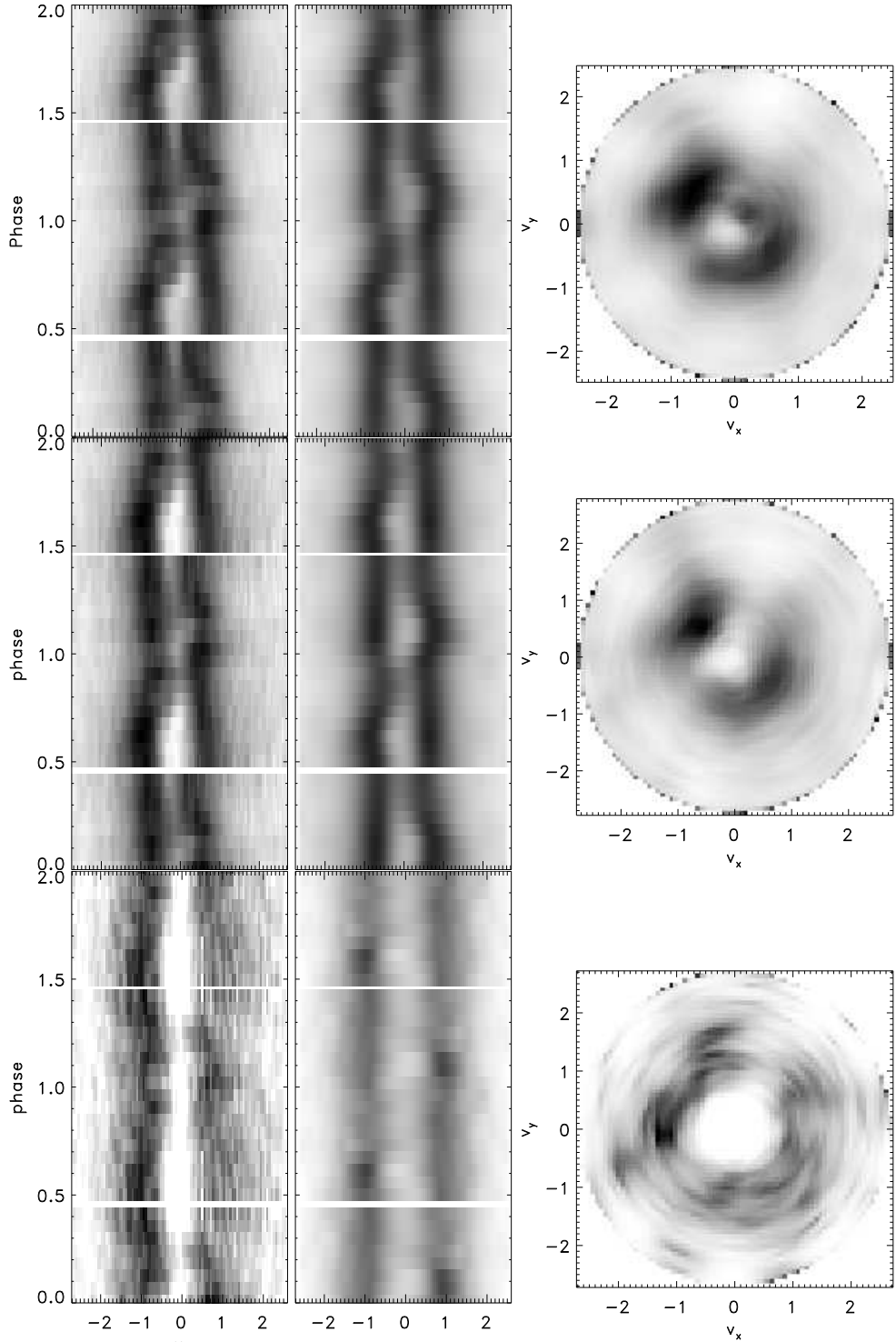}
\caption{\label{fig6}
Doppler maps for 1RXS\,J1808. From left to right: changes in the line profile with phase, changes in the line profile reconstructed from the Doppler map, and the Doppler map.
From top to bottom: the H$_\beta$, H$_\gamma$, and HeI$\,\lambda4471$ lines. }
\end{figure}

The binary was Doppler mapped using the $dopmap$
computer code developed by Spruit (1998). We
mapped the brightness distribution in the binary velocity
space in the H$_{\beta}$, H$_{\gamma}$,
H$_{\delta}$, HeI$~\lambda 4471$, HeII$~\lambda 4686$, HeI$~\lambda 5015$ lines. 
The maps of three lines are
presented in Fig. \ref{fig6}. The brightness distribution in
all three hydrogen lines has a similar structure with
two bright spots of different intensities. The brighter
spot with $V_{\rm x} \approx$ -600 km s$^{-1}$ and $V_{\rm y} \approx$ 400 km s$^{-1}$ is
near the accreting gas stream-accretion disk impact site.
 The second spot is more extended and lies on
the opposite side of the disk relative to the white
dwarf. The bright spots are much less distinct in the
maps from the neutral helium lines and are absent
altogether in the map corresponding to the ionized
helium line. This is indicative of a relatively low
temperature ($\approx$ 7000-9000 K) in the Balmer line
emission region. The constructed Doppler maps are
very similar to the brightness distribution maps for
IP Peg in quiescence (Neustroev et al. 2002), where
the two bright spots are interpreted as two-armed
spiral density waves.

As has already been noted in the Introduction, the
three-dimensional hydrodynamic modeling of accretion
disks performed by Bisikalo et al. (2001) predicts
the existence of a density perturbation in the disk
that rotates around the white dwarf with a period of
$\approx 0.1-0.2~ P_{\rm orb}$ and that is capable of producing the
binary brightness oscillations with the same period.
The second spot visible in the binary Doppler maps
can be associated with this density perturbation.

\section{Determination of binary parameters}

For cataclysmic variables with short orbital periods
 ($P_{\rm orb} <2^h$), there exists a fairly accurate semiempirical
 relation between the secondary mass and the
binary period. For 1RXS J1808, the mass-period
relations give $M_{\rm RD} \approx 0.12 M_{\odot}$
(Knigge 2006) and $M_{\rm RD} \approx 0.16 M_{\odot}$
(Howell et al. 2001). Therefore, we
can take the mass to be $M_{\rm RD} = 0.14 \pm 0.02 M_{\odot}$.

\begin{figure}
\centering
\includegraphics[angle=0,scale=1.2]{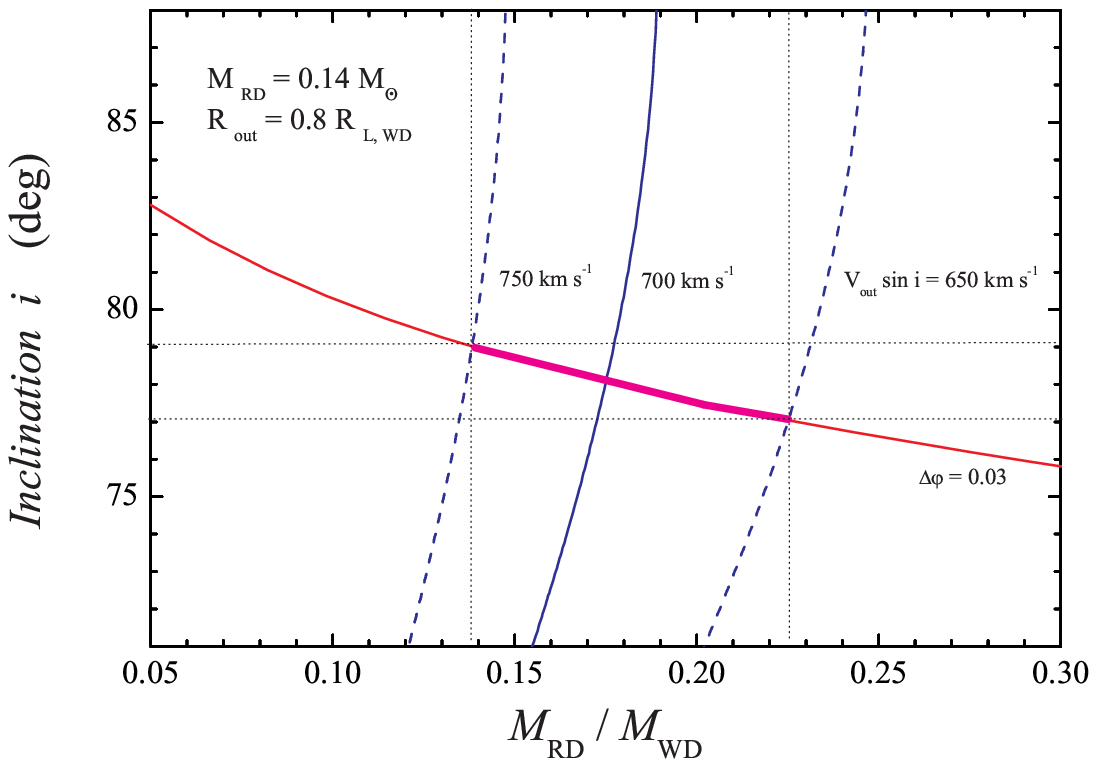}
\caption{\label{fig7}
Relations between the mass ratio $q$
and the binary inclination $i$
for the eclipse time $\Delta \varphi$\,=\,0.03
at various velocities of the outer disk edge. The secondary mass is $M_{\rm RD}  = 0.14 M_\odot$}
\end{figure}

The eclipse duration is related to the binary the relation between $i$ and $q = M_{\rm RD}/M_{\rm WD}$ (Horne 1985).
For our object, the observed eclipse
duration is approximately $\Delta \varphi$=0.03, which allows
the relation between $i$ and $q$ presented in Fig. \ref{fig7} to be
derived.

An additional relation between $i$ and $q$ is needed to
determine the parameters. Using the radial velocity
curve is undesirable because of the large initial phase
shift ($\approx$~0.25). Instead, we used the relation between $i$ 
and $q$ based on the assumption that the separation
between the peaks of double-humped emission lines
$\Delta\lambda$ corresponds to twice the projection of the Keplerian
 velocity at the outer disk edge onto the line of
sight:
\begin{equation} \label{u1}
     V_{\rm out} \sin i = \frac{\Delta\lambda}{2\lambda}\,c =
\sqrt{\frac{GM_{\rm RD}}{qR_{\rm out}}} \sin i.
\end{equation}

The outer disk radius is limited by the tidal interaction
 of the secondary star (Paczynski 1977) and it
is approximately 0.8 of the effective Roche lobe radius
$R_{\rm L,WD}$ (Eggleton 1983). The separation between the
emission line peaks depends weakly on the orbital
phase. Therefore, we used the separation between
the $H_{\beta}$ peaks at phase $\varphi \sim$ 0.5, when the hot spot
has the weakest effect on the line profile, which corresponds
 to the velocity at the disk edge $V_{\rm out}\sin i$ = 700 $\pm$ 50 km s$^{-1}$.
Using relation (\ref{u1}), we calculated
the domain of admissible values in the $i$--$q$ relation
(Fig. \ref{fig7}) The parameters $i$ and $q$ were determined from
it. As a result, we obtain $M_{\rm WD} = 0.8 \pm 0.22 M_{\odot}$, $i = 78^{\circ} \pm 1.^{\circ}5$
for $M_{\rm RD} = 0.14 \pm 0.02~ M_{\odot}$, $V_{\rm out} \sin i$ = 700 $\pm$ 50 km s$^{-1}$ and
$R_{\rm out} = 0.80 \pm 0.05~ R_{\rm L,WD}$.
We note, the calculated orbital velocity of the white
dwarf for these binary parameters is $K_1 \approx$ 70 km s$^{-1}$,
which closely corresponds to $K_1$ obtained from the radial velocity curve (see Table \ref{tbl-2}).

\section{Light curve modeling}

The R-band light curve of the binary being investigated
 (Fig. \ref{fig1}) exhibits quasi-periodic brightness oscillations
 with a period of .$\approx 0.^d$01331
($\approx 0.19~ P_{\rm orb}$). This
is confirmed by our analysis of the power spectrum
(Fig. \ref{fig2}) computed with the EFFECT code (V. Goransky,
 the Sternberg Astronomical Institute, Moscow).
The corresponding brightness oscillations are not observed
 in the V band light curves (Fig. \ref{fig1}) most likely
because of the poorer quality of the observational data,
 but a significant change in the depths of neighboring
 eclipses was found in the August 12 observations.

To explain the period found in the R band and
the change in the depth of the August 12 eclipse
in the V band, we modeled these light curves using
a modified Magnitude code written by M. Stupalov
(Shimanskii et al. 2002).

The model included a secondary, a white dwarf
with a surface temperature $T$ = 50000 K, and a standard
 accretion disk (Shakura and Sunyaev 1973) with
an external hot line formed from the stream of the secondary
 and a spot revolving around the white dwarf
with a period $P_{\rm sp2}\approx 0.154~ P_{\rm orb}$ and corresponding to
a density wave (see the Section "Analysis of the Spectroscopic
 Observations"). The second spot is needed
to describe the quasi-periodic brightness oscillations
and the different depths of neighboring eclipses. The
revolution period of the spot differs from the observed
period of the binary brightness oscillations ($P_{\rm qpo} \approx 0.2 P_{\rm orb}$),
because the brightness oscillation period is
the beat period between the revolution period of the
spot around the WD and the orbital period of the
binary $1/P_{\rm qpo} = 1/P_{\rm sp2} - 1/P_{\rm orb}$.

\begin{figure}
\centering
\includegraphics[clip,width=130mm]{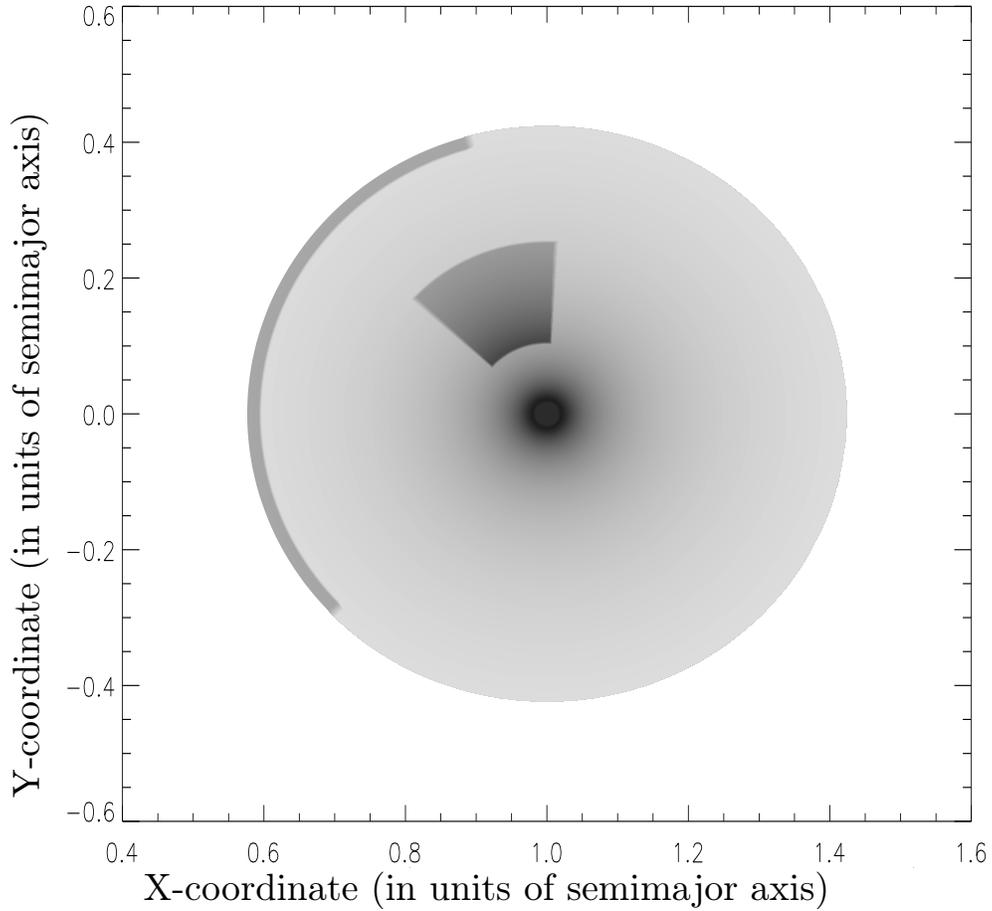}
\caption{\label{fig8}
The model of an accretion disk with two spots. The red dwarf is on the left, at $X<0.4$.
The brightness of the disk regions corresponds to the temperature. The hot line from the stream of the secondary is shown schematically (in the model,
it occupies the disk side) with an artificially enhanced brightness for clarity. The second spot revolves around the disk center and, consequently, can be at different angular positions. }
\end{figure}

\begin{figure}
\centering
\includegraphics[clip,angle=0,width=130mm]{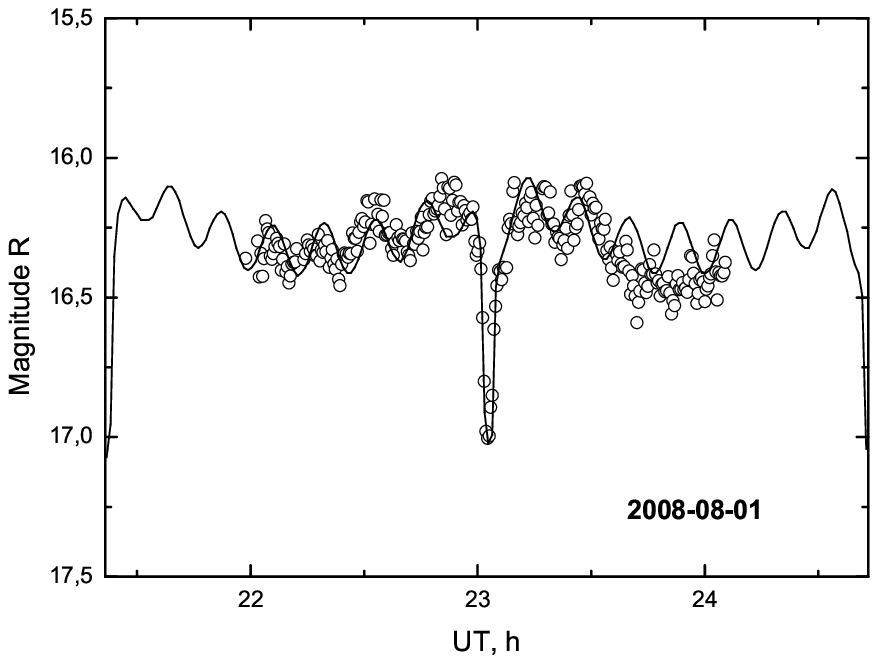}
\includegraphics[clip,angle=0,width=130mm]{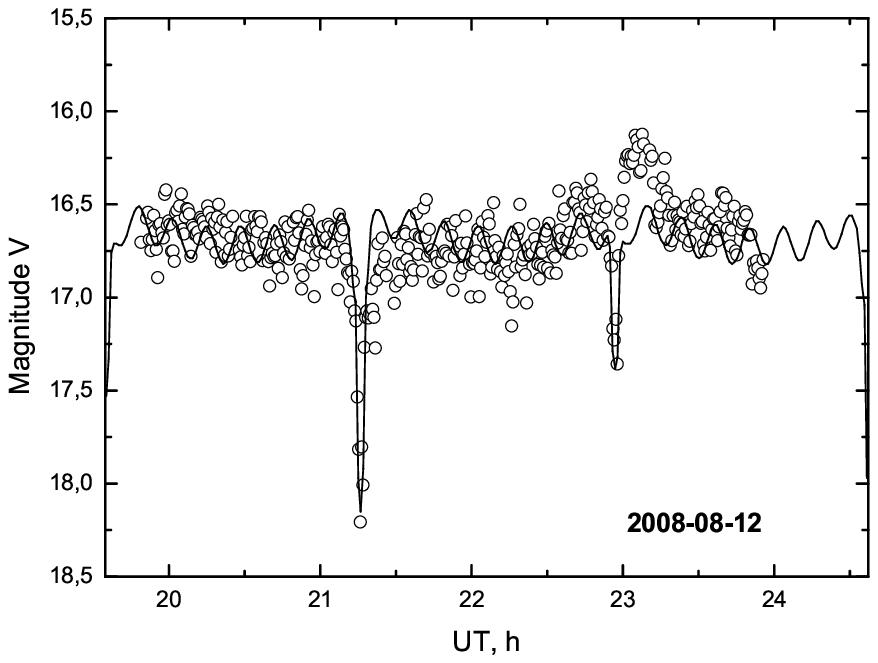}
\caption{\label{fig9}
Model (a) R-and (b) V-band light curves.
}
\end{figure}

The essence of our modeling consists in dividing
 the surface of objects into surface elements and
calculating the radiation intensity in the blackbody
approximation, the visibility condition, and the possibility
 of an eclipse for each surface element. To
calculate the total flux from the binary, we add up
the fluxes from all surface elements visible at a given
phase. The binary parameters found above were used
in the model. The best fit is achieved at an accretion
rate $\dot M \approx 6.3 (\pm 3) \times 10^{15}$ g s$^{-1}$  ($\chi^2_{\rm d.o.f.} \approx 3$). The
positions of the spots are shown in Fig. 8. The first
spot with $T_1$ = 5200 K corresponding to the stream.
disk impact site (hot line) is located near the outer
disk radius, has the relative width $\Delta R_1$=0.017 $R_{\rm out}$,
and lies between the azimuth angles $\Phi$ $105^{\circ}-225^{\circ}$.
The azimuth angle is measured from the straight line
connecting the centers of mass of the WD and the
secondary. The direction opposite to the direction to
the WD was taken as zero angle. The angles in Fig. \ref{fig8}
are measured counterclockwise. The second spot has
the opening angle $\Delta \Phi_2=50^{\circ}$
and is bounded by the
radii of 0.22 and 0.58 $R_{\rm out}$. Here, $R_{\rm out}$ is the outer
disk radius. The brightness temperature in the spot
is almost a factor of 2 higher than that in the disk and
changes from 18000 to 10000 K. It should be noted
that the parameters of the second spot are not rigidly
determined and admit concurrent temperature and
size variations in fairly wide ranges (approximately by
a factor of 2) without any noticeable deterioration of
the accuracy of the fit to the observed light curves.
For example, the azimuthal extent of the second spot
can change from 25 to 100 degrees with the corresponding
 change in mean temperature from 16000 to
9700 K.
The modeling results are presented in Fig. \ref{fig9}. The
proposed model is capable of explaining the different
 depths of neighboring eclipses in the V band
and describing some of the quasi-periodic oscillations
 in the R band. In reality, the disk apparently
has a more complex structure changing with time.
For example, the V-band observations suggest that
the hot spot clearly has different intensities at different
 times, while the R-band observations reveal no
quasi-periodic brightness oscillations after phase 0.3 (23.5 h).

\section{Discussion}

Let us consider in more detail the final model.
The derived accretion rate $\dot M = 6.3 \times 10^{15}$ g/s $\approx 10^{-10} M_\odot/$year
corresponds to the expected accretion
rate from a CV with this period (Howell et al. 2001).
The estimated effective temperature at the outer
disk edge $T_{\rm out} \approx 3400$~K at $R_{\rm out} \approx 2.1 \times 10^{10}$~cm is
lower than that required for a steady-state accretion
 disk in outburst ($T_{\rm out} \approx 5500$~K, Cannizzo and
Wheeler 1984). This may imply that the component
mass ratio $q$ can be higher than the mean value that
we adopted in our modeling and it is closer to 0.22.
The outer disk radius may also be smaller than the
value of 0.8 $R_{\rm L,WD}$ we adopted.

Let us investigate the possible physical nature
of the radiation from the second spot. If the derived
 color temperature of the spot is assumed to
be close to the effective one, then its bolometric luminosity
 can be estimated. Its value ($\approx (1-3) \times 10^{32}$ erg s$^{-1}$)
is approximately of the same order of
magnitude as the bolometric luminosity of the disk
$L_{\rm d} = GM_{\rm WD} \dot M/(2R_{\rm WD})$\
$\approx 5 \times 10^{32}$ erg s$^{-1}$
and exceeds the luminosity of the same area of the disk without
any spot by a factor of 7-16. No additional energy
dissipation in the density wave is apparently capable
of providing such a luminosity. External irradiation by
the central regions of the disk and WD can be another
source of energy for the radiation from the second
spot. However, the irradiating flux in cataclysmic
variables is usually lower than the intrinsic flux in the
disk. Let us make simple estimates. The intrinsic flux
in a standard steady-state disk is
\begin{equation} \label{ud1}
   F_0 \approx \frac{3}{8\pi} \frac{G \dot M M_{\rm WD}}{R^3} = \frac{3}{4\pi}
   \frac{L_{\rm d}}{R^2}\frac{R_{\rm WD}}{R},
\end{equation}
the external flux from the central parts of the disk can
be estimated as (Shakura and Sunyaev 1973)
\begin{equation} \label{ud2}
   F_{\rm irr} \approx \frac{1}{8}\frac{L_{\rm d}}{2\pi R^2} \left(\frac{z(R)}{R}\right)^2,
\end{equation}
and the irradiating flux from an isotropic central
source of luminosity $L_{\rm d}$ located at height $h$ above the
disk plane is expressed as
\begin{equation} \label{ud3}
   F_{\rm irr} \approx \frac{L_{\rm d}}{4\pi R^2}
   \left(\frac{h}{R}+\frac{1}{8}\frac{z(R)}{R}\right).
\end{equation}
Here, $z(R)$
is the disk half-thickness at radius $R$. As
a result, the ratio of the irradiating flux to the intrinsic
one is
\begin{equation} \label{ud4}
   \frac{F_{\rm irr}}{F_0} \approx \frac{1}{3}\frac{h}{R_{\rm WD}} +
   \frac{1}{24}\frac{z(R)}{R_{\rm WD}} +\frac{1}{12}\frac{z(R)}{R_{\rm WD}}\frac{z(R)}{R}.
\end{equation}

The above formulas are valid for geometrically thin
($z(R)/R \le 0.1$) steady-state $\alpha$-disks (Shakura and
Sunyaev 1973). It is also assumed that the luminosity
 of the central source is provided by accretion
(a boundary layer or an accretion column) and its
maximum possible luminosity is equal to the disk
luminosity. Accordingly, the height of the central
source above the disk plane cannot exceed $R_{\rm WD}$.

According to the constructed model, $R_{\rm WD} \approx 7 \times 10^8$ cm,
$R_{\rm sp2} = 0.4 - 1 \times 10^{10}$ cm, and $z(R_{\rm sp2}) \le 0.4 - 1 \times 10^{9}$
cm (see, e.g., Suleimanov et al. 2007).
This means that the disk half-thickness in the region
of the second spot, even if the probable disk thickening
 is taken into account, is comparable to the
WD radius and the external irradiation flux cannot
exceed considerably the intrinsic flux from the disk.
Strong irradiation of the disk is possible only in
binaries with neutron stars whose radii ($\approx 10^6$cm) are
much smaller than those of white dwarfs at the same
geometrical sizes of the disks. Given that not all of
the externally incident flux is thermalized, it may be
ultimately concluded that external irradiation cannot
provide the radiation of the second spot.

This conclusion is valid only if the luminosity of
the second spot was determined correctly. If, alternatively,
 its radiation is not thermalized and exceeds the
disk luminosity by several tens of percent only in the
optical spectral range (for example, due to emission
lines), then external irradiation is, in principle, capable
 of providing the observed optical variability, just as
in intermediate polars.

Consider the possibility that 1RXS J1808 is an
intermediate polar. Several objects of this type, for
example, DQ Her (Walker 1956; Saito and Baptista
 2009) and LS Peg (Rodrigez-Gil et al. 2001),
exhibit variability both with the WD spin period and
with the beat period between the orbital and spin
periods. The amplitude of the optical brightness oscillations
 with the WD spin period is low ($\approx 0.^m1$ for DQ Her and $\approx 0.^m003$
for LS Peg), while the detection
of the beat period produced by the reflected radiation
is most significant in the hydrogen emission lines and
HeII\,4686.

The HeII\,4686 line in emission is formed mainly
through photoionization by external hard radiation
and is a good indicator of external disk irradiation.
 For example, in the spectrum of DQ Her
(Chanan et al. 1978), the flux in HeII\,4686 is comparable
 to that in the Balmer lines. Therefore, the
 weakness of the HeII\,4686 line and the Bowen blend
in the spectrum of the source being investigated
serves as direct evidence for insignificant external
irradiation of the disk in 1RXS J1808. This is also
suggested by the absence of any structure in the
Doppler tomogram of the HeII\,4686 line.

If the source of additional optical radiation that we
consider to be the second spot were associated with
the rotating WD, then the eclipses would be deeper
against the background of an enhanced brightness of
the binary outside eclipse, whereas the reverse is true
-- the eclipses are deeper at a general reduction in
the binary brightness (see Fig. \ref{fig1}), which is consistent
with the model of a second spot on the disk presented
here.
On the whole, the central regions of the disk near
the WD are the main eclipsed source of optical radiation.
 The WD itself, even if very hot (we took a
temperature of 50000 K), makes a minor contribution
to the total brightness of the binary. We modeled the
binary light curves by assuming the inner disk radius
to exceed the WD radius. Thus, we simulated the
appearance of a magnetosphere around a magnetized
WD. Satisfactory results can be achieved only at
an inner disk radius that does not exceed 3--4 WD
radii ($R_{\rm in} < 3 \cdot 10^9 \approx 0.15 R_{\rm d}$~ cm). At the accretion
rate under consideration (even when its increase by
several times as the inner disk radius increases is
taken into account), the WD magnetic field strength
turns out to be less than 10$^5$ G. Taking into account
all of the above arguments, we may conclude that
1RXS J1808 is not an intermediate polar with a high
probability.

The only remaining possibility is the release of
energy in the second spot through direct accretion
apart from the disk (Hellier 1993, see also Neustroev
et al. 2011). This requires assuming that part of the
accreting material moves under the accretion disk
and is stopped by a thickening on it in the form of a
tidal density wave. If the mass accretion rate above
the disk is comparable to that in the disk, then the
interaction of the accreting stream with the density
wave located at a distance of 5--6 WD radii provides
the required bolometric luminosity of the spot. The
possibility that the presumed stream of matter interacts
 not with the density wave but with the WD
magnetosphere must not be ruled out. However, in
this case, to explain the spot revolution around the
WD, we have to assume that the magnetosphere is
highly asymmetric. The hypothesis of a tidal wave
is also supported by the coincidence of the revolution
period of the second spot with the theoretically predicted
 value (Bisikalo et al. 2001).

\section{Conclusions}

We showed that 1RXS J1808 is a cataclysmic
variable with a nonuniform brightness distribution in
the disk with the component masses 
$M_{\rm WD} = 0.8 \pm 0.22 M_{\odot}$, $M_{\rm RD} = 0.14 \pm 0.02~ M_{\odot}$
and the binary
inclination $i = 78^{\circ} \pm 1.^{\circ}5$.
Our Doppler mapping and
light curve modeling revealed at least two spots in the
accretion disk, at the stream.disk impact site, and a
spot corresponding to a density wave and revolving
around the white dwarf with a period of .$\approx 0.154~ P_{\rm orb}$.
The luminosity of the second spot is most likely related
 to direct accretion of matter onto the disk thickening
 produced by a tidal density wave.

This binary is most likely a cataclysmic variable
star in outburst similar in properties to nova-like
stars, and further photometric and spectroscopic observations
 are required to confirm this conclusion.

We sincerely thank the Big Telescope Committee
 of the Special Astrophysical Observatory of the
Russian Academy of Sciences for the long-term support
 of our programs of research on the spectra of
close binary systems. We are also grateful to the
TUBITAK for partial support in using RTT-150 (the
Russian-Turkish 1.5-m telescope in Antalya). We
thank A.I. Galeev for help in the photometric observations
 with RTT-150. This work was supported by
the Russian Foundation for Basic Research (project
no. $09$-$02$-$97013$ r-povolzhie-a)

\end{document}